\documentclass[aps,prl,twocolumn,showpacs,superscriptaddress]{revtex4-2}
\usepackage{graphicx,amsmath,amssymb,amsfonts}
\usepackage[latin1]{inputenc}
\usepackage{array}
\usepackage{multirow}
\usepackage{float}
\usepackage{color}
\usepackage[normalem]{ulem}
\usepackage{booktabs}

\usepackage{hyperref}
\hypersetup{
	colorlinks = true,
	linkcolor = [rgb]{0.70, 0.13, 0.13},
	citecolor = [rgb]{0.25, 0.41, 0.88},
	urlcolor = [rgb]{0.25, 0.41, 0.88}
}
\usepackage{pifont}

\begin{document}

\title{Structural and magnetic characterization of CeTa$_7$O$_{19}$ and YbTa$_7$O$_{19}$ with two-dimensional pseudospin-1/2 triangular lattice}

\author{Feihao Pan}
\author{Songnan Sun}
\affiliation{Laboratory for Neutron Scattering and Beijing Key Laboratory of Optoelectronic Functional Materials and MicroNano Devices, Department of Physics, Renmin University of China, Beijing 100872, China}
\affiliation{Key Laboratory of Quantum State Construction and Manipulation (Ministry of Education), Renmin University of China, Beijing, 100872, China}
\author{Alexander I. Kolesnikov}
\author{Matthew B. Stone}
\affiliation{Neutron Scattering Division, Oak Ridge National Laboratory, Oak Ridge, Tennessee 37831, USA}
\author{Jiale Huang}
\author{Daye Xu}
\author{Chenglin Shang}
\author{Bingxian Shi}
\author{Xuejuan Gui}
\author{Zhongcen Sun}
\affiliation{Laboratory for Neutron Scattering and Beijing Key Laboratory of Optoelectronic Functional Materials and MicroNano Devices, Department of Physics, Renmin University of China, Beijing 100872, China}
\affiliation{Key Laboratory of Quantum State Construction and Manipulation (Ministry of Education), Renmin University of China, Beijing, 100872, China}

\author{Jinchen Wang}
\author{Juanjuan Liu}
\author{Hongxia Zhang}
\author{Zhengxin Liu}
\author{Peng Cheng}
\email[Corresponding author: ]{pcheng@ruc.edu.cn}
\affiliation{Laboratory for Neutron Scattering and Beijing Key Laboratory of Optoelectronic Functional Materials and MicroNano Devices, Department of Physics, Renmin University of China, Beijing 100872, China}
\affiliation{Key Laboratory of Quantum State Construction and Manipulation (Ministry of Education), Renmin University of China, Beijing, 100872, China}

\begin{abstract}
Triangular lattice antiferromagnets are prototypes for frustrated magnetism and may potentially realize novel quantum magnetic states such as a quantum spin liquid ground state. A recent work suggests NdTa$_7$O$_{19}$ with rare-earth triangular lattice is a quantum spin liquid candidate and highlights the large family of rare-earth heptatantalates as a framework for quantum magnetism investigation. In this paper, we report the structural and magnetic characterization of CeTa$_7$O$_{19}$ and YbTa$_7$O$_{19}$. Both compounds are isostructural to NdTa$_7$O$_{19}$ with no detectable structural disorder. For CeTa$_7$O$_{19}$, the crystal field energy levels and parameters are determined by inelastic neutron scattering measurements. Based on the crystal field result, the magnetic susceptibility data could be well fitted and explained, which reveals that CeTa$_7$O$_{19}$ is a highly anisotropic Ising triangular-lattice antiferromagnet ($g_z$/$g_{xy}$$\sim$3) with very weak exchange interaction (J$\sim$0.22~K). For YbTa$_7$O$_{19}$, millimeter sized single crystals could be grown. The anisotropic magnetization and electron spin resonance data show that YbTa$_7$O$_{19}$ has a contrasting in-plane magnetic anisotropy with $g_z$/$g_{xy}$$\sim$0.67 similar as that of YbMgGaO$_4$. The above results indicate that CeTa$_7$O$_{19}$ and YbTa$_7$O$_{19}$ with pseudospin-1/2 ground states might either be quantum spin liquid candidate materials or find applications in adiabatic demagnetization refrigeration due to the weak exchange interaction.

\end{abstract}

\maketitle

\section{Introduction}
The geometrically frustrated triangular-lattice antiferromagnets (TLAF) that carry spin-1/2 have attracted great attentions in recent decades. Due to the spin frustration and strong quantum fluctuations, they may host the famous quantum spin liquid state (QSL)\cite{RMP,collin}. QSLs feature long-range quantum entanglement and support fractionalized excitation. They are also an important material basis for the realization of topological quantum computation in the future. Besides, TLAFs are also proposed to host other exotic quantum states of matter, such as the spin supersolid which may generate a giant magnetocaloric effect and be quite useful in adiabatic demagnetization refrigeration\cite{Sugang2024,SJM}.

For Kramer rare-earth (RE) ions such as Ce, Nd and Yb, if they have a well separated ground-state doublet due to the crystal field effect, they can be viewed to carry an effective spin $S$=1/2 at low temperatures. In recent years, RE-based TLAFs have become important resources in searching for QSL candidates. The well known examples include the YbMgGaO$_4$ and AYbX$_2$ (A=Cs,K,Na; X=S,O,Se) family of materials\cite{YMGO, NRX27, NRX28, NRX29, NRX30, NRX31, NRX32, NRX33, NRX34, NRX35, NRX36, NRX37}. However, an unambiguous confirmation of a QSL in these TLAFs is still lacking. There are arguments that many experimental evidences for QSLs such as spin excitation continuum or lack of magnetic order at very low temperature could also be caused by spin glass or random singlet phases\cite{QSL1, QSL2}. It has been known that sizeable structural disorder may induce mimicry of a QSL\cite{YMGO1, YMGO_WEN}. Therefore searching for and characterizing more TLAFs without atomic-site disorder would help to clarify many on-going questions in this research field.  

NdTa$_7$O$_{19}$ is recently proposed as a QSL candidate\cite{NM2022}. In this compound, the Nd$^{3+}$ form a two-dimensional (2D) triangular lattice without structural disorder. Inelastic neutron scattering (INS) measurements on the crystal-electric-field (CEF) levels confirm its magnetic ground state is characterized by effective spin-1/2 degrees of freedom. Combined with magnetization, electron spin resonance (ESR) and muon spin relaxation spectroscopy results, NdTa$_7$O$_{19}$ is identified as a rare example with both highly anisotropic Ising-like exchange interactions and QSL features. Thereafter, the synthesis and structure of RETa$_7$O$_{19}$ (RE= Pr, Sm, Eu, Gd, Dy, Ho) with triangular lattice were reported\cite{TZM}. However, there are still two possible members in this material family, namely CeTa$_7$O$_{19}$ and YbTa$_7$O$_{19}$, whose structural and magnetic properties are still unknown.

Based on this motivation, we have successfully synthesized CeTa$_7$O$_{19}$ and YbTa$_7$O$_{19}$. They are found to be isostructural to NdTa$_7$O$_{19}$ with a disorder-free 2D triangular lattice. Magnetic susceptibility, inelastic neutron scattering and ESR measurements reveal that CeTa$_7$O$_{19}$ has a strong Ising-like anisotropy with effective spin-1/2 while YbTa$_7$O$_{19}$ exhibits an in-plane magnetic anisotropy. Both CeTa$_7$O$_{19}$ and YbTa$_7$O$_{19}$ have very weak antiferromagnetic interactions, their magnetic properties are discussed with possible connections with a QSL ground state.

\begin{figure}
	\includegraphics[width=7.8cm]{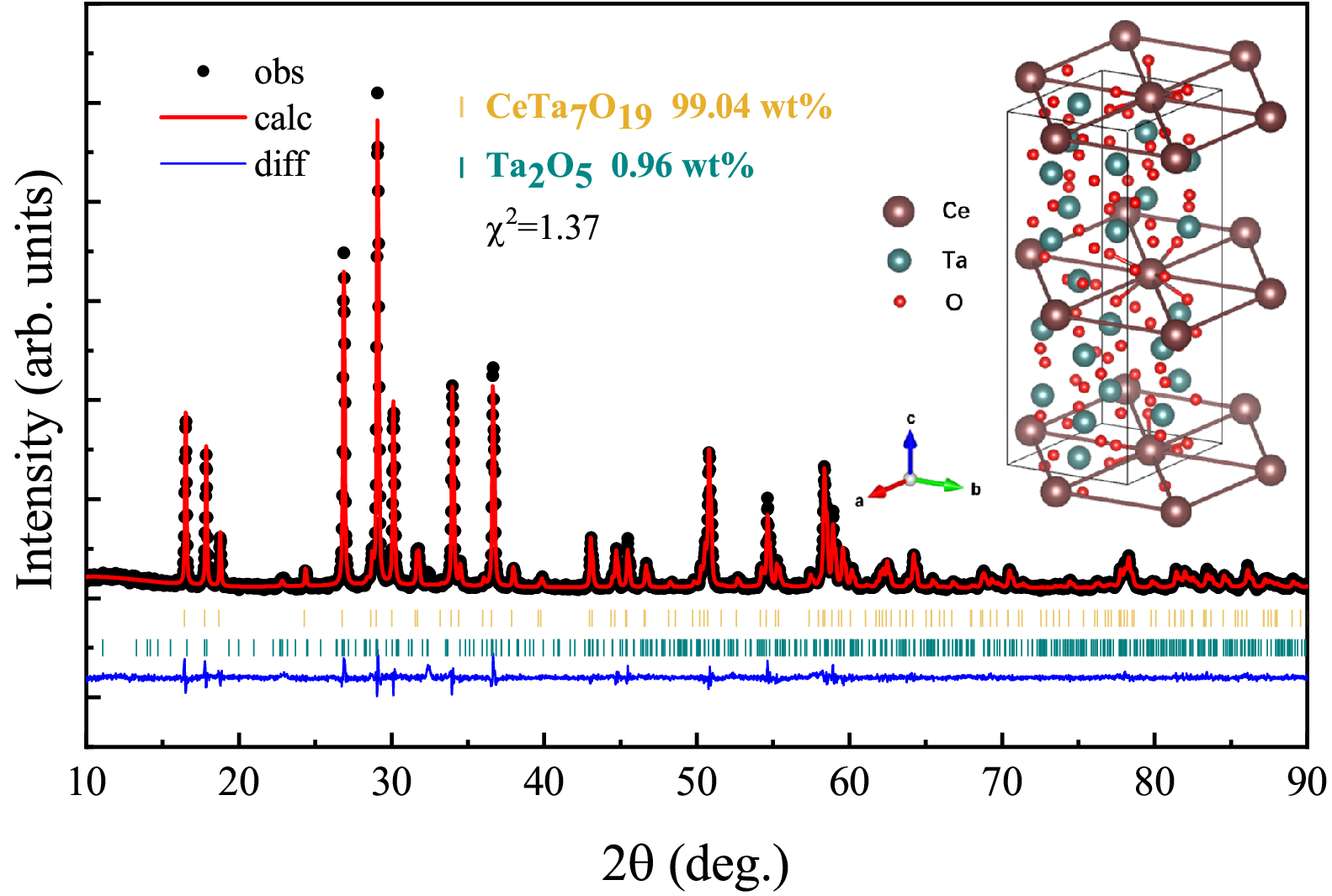}
	\caption {X-ray diffraction patterns and Rietveld refinement on CeTa$_7$O$_{19}$ powders. The inset shows the crystal structure of CeTa$_7$O$_{19}$.} \label{Fig1}
\end{figure}

\begin{figure*}[htbp]
	\centering
	\includegraphics[width=\textwidth]{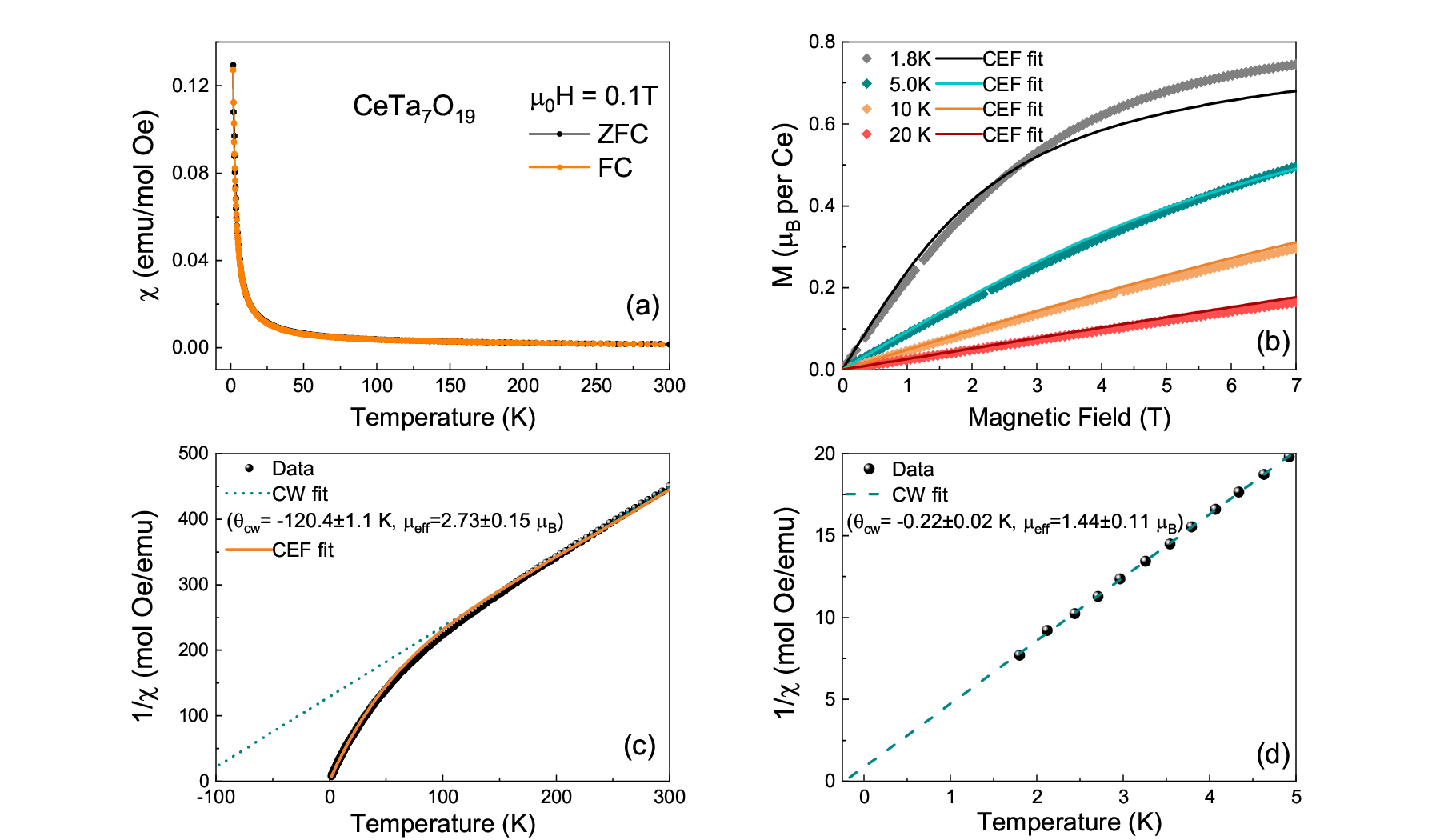}
	\caption {(a) Temperature-dependent magnetic susceptibility $\chi$(T) of CeTa$_7$O$_{19}$. (b) Isothermal magnetization M(T) of CeTa$_7$O$_{19}$ at selected temperatures and the solid lines are corresponding calculated result based on the crystal field parameters obtained from neutron scattering. (c) The Curie-Weiss fit on the high temperature susceptibility data (dotted line) and the crystal field model fit of the susceptibility data (solid line). (d) The Curie-Weiss fit on the susceptibility data below 5~K.} \label{Fig2}
\end{figure*}

\section{methods}

Polycrystalline CeTa$_{7}$O$_{19}$ samples were prepared via the solid-state reaction method. Initially, CeO$_{2}$ powder and Ta$_{2}$O$_{5}$ powder were mixed in a ratio of 1:3.5, thoroughly ground and pressed into a pellet. Subsequently, the mixture was placed in an alumina crucible, heated at 950°C for 7 days in a muffle furnace and finally furnace cooled to room temperature. The product at this stage contains large amount of impurity phases. In order to get phase-pure CeTa$_{7}$O$_{19}$ powder, the product was re-ground, pressed and heated at 1000°C in the furnace for 5 days. This procedure was repeated with the heating temperature increased by 50°C in each firing. Then after the final reaction at 1150°C for 5 days, nearly phase-pure CeTa$_{7}$O$_{19}$ polycrystalline samples which exhibit a light yellow color could be obtained.

For the growth of YbTa$_{7}$O$_{19}$ single crystals, firstly the precursor YbOCl powder needs to be prepared by the solid-state reaction of NH$_4$Cl and Yb$_2$O$_3$ powder, similar as that in preparing DyOCl in our previous work\cite{DyOCl}. Then YbOCl, Ta$_{2}$O$_{5}$, and Ta were mixed in a molar ratio of 2:1:1, ground and pressed into pellet. The mixture was sealed in a quartz tube and evacuated to a high vacuum of 5.0$\times$10$^{-2}$ Pa. Subsequently, the quartz tube was heated to 950°C at a rate of 50°C/h and maintained at this temperature for 7 days. After furnace cooling to room temperature, the hexagonal plate-like YbTa$_{7}$O$_{19}$ single crystals with dark brown color and typical dimensions of 1.5$\times$1.5$\times$0.1 mm$^{3}$ could be found from the product. 

We should mention that phase-pure polycrystalline YbTa$_7$O$_{19}$ could not be obtained from direct solid-state reaction method. The impurity phases always exist with large volume fraction. Therefore the data on polycrystalline YbTa$_7$O$_{19}$ were collected on powders crushed from single crystals. Alternatively, it is found that single crystals of CeTa$_{7}$O$_{19}$ could not be grown using similar method described above.

X-ray diffraction (XRD) patterns of the samples were collected from a Bruker D8 Advance X-ray diffractometer using Cu K$_{\alpha}$ radiation. Magnetization measurements were carried out in Quantum Design MPMS3. Inelastic neutron scattering experiments were performed on the fine-resolution Fermi chopper spectrometer SEQUOIA at the Spallation Neutron Source of Oak Ridge National Laboratory and employed neutrons with incident energies (E$_i$) of 140~meV\cite{Granroth_2010,stone}. 4.6~g CeTa$_7$O$_{19}$ and LaTa$_7$O$_{19}$ powder samples were measured at 6.5~K. The data for non-magnetic analogue LaTa$_7$O$_{19}$ served as background, which helps us to subtract phonon contributions from the measured spectra. 

The ESR measurement of the powder sample was performed on the X-band ESR spectrometer (CIQTEK, EPR-100) with a dry cooling system. The spectra were recorded with the microwave power of 0.2~mW at the frequency of 9.77~GHz, modulation amplitude of 1~Gauss, time constant of 0.1~s.

\section{Results and discussions}

\subsection{CeTa$_7$O$_{19}$}

According to the current inorganic crystal structure database (ICSD), CeTa$_7$O$_{19}$ forms a crystal structure with honeycomb lattice of Ce and there is no record for YbTa$_7$O$_{19}$. However, as shown in Fig. 1, the Rietveld refinement result on the powder XRD patterns of CeTa$_7$O$_{19}$ reveals that it is isostructural to NdTa$_7$O$_{19}$ with Ce$^{3+}$ ions forming the 2D triangular lattice. The detailed crystallographic data are shown in the Supplemental Material\cite{supp}. There is some slight Ta$_2$O$_5$ impurity identified from XRD which would not affect the following physical properties measurements. Within the $ab$-plane, the nearest Ce-Ce distance is 6.230~\AA. This value is close to the nearest Nd-Nd distance 6.224~\AA~in NdTa$_7$O$_{19}$\cite{NM2022}.

Fig. 2(a) presents the magnetic susceptibility $\chi(T)$ data of CeTa$_7$O$_{19}$. Both the zero-field-cooling and field-cooling data overlaps well and show no signs of any magnetic order down to 1.8~K. From the temperature-dependent inverse susceptibility shown in Fig. 2(c), the data at above 120~K have linear temperature dependence which means that it could be well fitted by the Curie-Weiss (CW) model. The CW fit gives very large values of CW temperature $\theta_{CW}$ and effective moment $\mu_{eff}$ which should be notably affected by the crystal field excitation. Below 120~K, the 1/$\chi$(T) data gradually deviate from the linear temperature dependence. However an additional CW behavior restores below 5~K as shown in Fig. 2(d) and the fit yields $\theta_{CW}$=-0.22~K and $\mu_{eff}$=1.44~$\mu_{B}$. For rare-earth compounds with very weak exchange interaction, such low temperature range is usually far above exchange coupling and far below the first CEF excitation. Therefore the CW fitting results on this range may reflect the strength of antiferromagnetic interaction and the pesudospin-1/2 state in CeTa$_7$O$_{19}$, as in the case of NdTa$_7$O$_{19}$\cite{NM2022} and Ce$_2$Zr$_2$O$_7$\cite{Gao2019}. We will make further discussions after considering the following crystal field results.

For rare-earth compounds, the experimental determination of CEF levels and parameters is important to understand their magnetism. For CeTa$_7$O$_{19}$, Ce$^{3+}$ 
with $J$=5/2 has an odd number of $f$ electrons and the CEF potential from oxygen will split them into three Kramers doublets. Therefore at low temperature, two CEF excitations from the ground state to the two excited states are expected to be observed in the inelastic neutron scattering (INS) data. From the INS spectra of CeTa$_7$O$_{19}$ at 6.5~K in Fig. 3(a), these two excitations can be identified at low-$Q$ range while the lattice (phonon) excitations dominate in the high-$Q$ range. After subtracting the spectra of the non-magnetic analogue LaTa$_7$O$_{19}$, two flat CEF excitation bands are clearly observed as shown in Fig. 3(b). It is known that the CEF excitations are not coupled to propagating modes and do not possess a characteristic dispersion. Their scattering intensities decrease with $Q$ following the magnetic form factor $F(Q)^2$, which is consistent with experimental result shown in the inset of Fig. 3(c). In Fig. 3(c), the fit on the data summed over 2.5~\text{\AA}$^{-1}$$\leqslant$ $Q$ $\leqslant$ 4.5~\text{\AA}$^{-1}$ could confirm the energy of two CEF exciations are located at 42.9~meV and 67.1~meV. Therefore the schematic diagram of the CEF energy levels in CeTa$_7$O$_{19}$ is presented in Fig. 3(d).

\begin{figure*}[htbp]
	\centering
	\includegraphics[width=\textwidth]{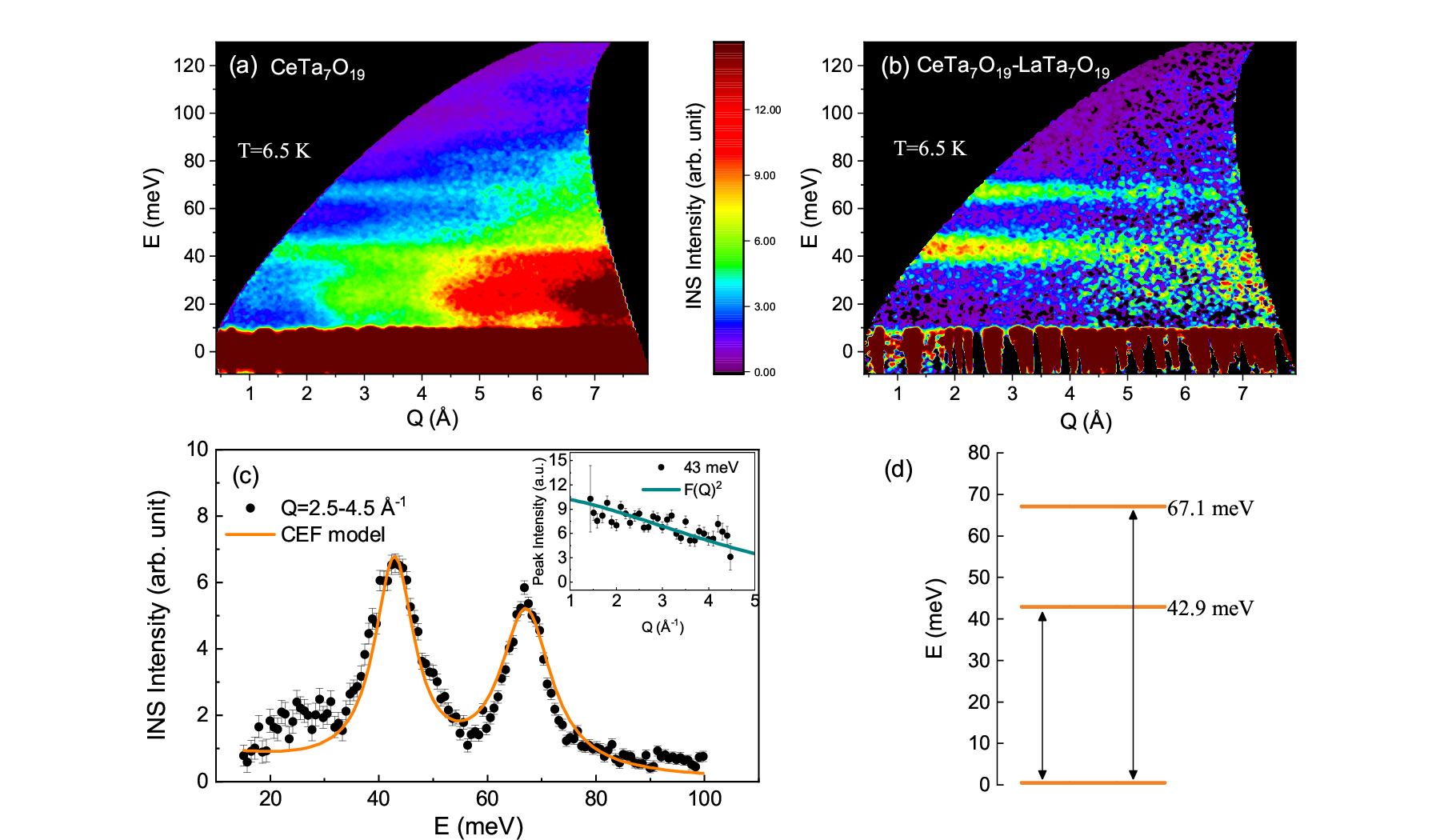}
	\caption {(a) INS spectra from CeTa$_7$O$_{19}$ powder sample with an incident neutron energy E$_i$=140~meV. (b) INS spectrum after subtracting phonon contributions and other background scattering by measuring non-magnetic analogue LaTa$_7$O$_{19}$. (c) Energy spectrum obtained by integrating the INS data in (b) over the ranges 2.5~\text{\AA}$^{-1}$-4.5~\text{\AA}$^{-1}$. The solid line is the CEF model fit and the inset shows the $Q$-dependent intensity of 43~meV peak agrees well with the magnetic form factor. (d) A scheme of the CEF splitting of the Ce$^{3+}$ ground multiplet is shown.} \label{Fig3}
\end{figure*}

The CEF parameters could also be obtained through the fit of INS data. For CeTa$_7$O$_{19}$, the Ce$^{3+}$ has a $D_{3v}$ local point-group symmetry. Then the CEF Hamiltonian can be written as $H_{CEF}=B_2^0\hat{O}_2^0+B_4^0\hat{O}_4^0+B_4^3\hat{O}_4^3+B_6^0\hat{O}_6^0+B_6^3\hat{O}_6^3+B_6^6\hat{O}_6^6$, where the $B_l^m$ are the CEF parameters and the $\hat{O}_l^m$ are the CEF Stevens equivalent operators. Since Ce$^{3+}$ has $J$=5/2 and all the sixth-order terms are zero, only three CEF parameters need to be derived from the fit of INS spectra. Using Mantid software\cite{Mantid}, the best fit achieves with $B_2^0=-1.309$, $B_4^0=-0.020$ and $B_4^3=3.335$ (unit: meV). Then the eigenstates of the CEF Hamiltonian could be calculated and shown in Table.\ref{T1} in the $|m_J\rangle$ basis.

\begin{table}\fontsize{5pt}{9pt}\selectfont
\resizebox{6cm}{!}{
\begin{tabular}{c|c|c|c} \hline 
	$\left| \pm m_J \right\rangle $ & $\pm \omega_0$ & $\pm \omega_1$ & $\pm \omega_2$ \\ \hline
	$\left| \pm 5/2 \right\rangle $ &0 &0 &0 \\
	$\left| \pm 3/2 \right\rangle $ &0 &1 &0\\
	$\left| \pm 1/2 \right\rangle $ &$\pm$0.577 &0 &0.817\\
	$\left| \mp 1/2 \right\rangle $ &0 &0 &0\\
	$\left| \mp 3/2 \right\rangle $ &0 &0 &0\\
	$\left| \mp 5/2 \right\rangle $ &0.817 &0 &$\mp$0.577\\ \hline
	E(meV) &0 &42.9 &67.1 \\ \hline
\end{tabular}
}
\caption{The CEF eigenstates wave functions given in the 	$\left| \pm m_J \right\rangle $ basis and the corresponding energies of three Ce$^{3+}$ doublets in CeTa$_7$O$_{19}$.}
\label{T1}
\end{table}

The CEF parameters determines the rare-earth single-ion magnetic anisotropy. For Ce-based magnetic materials, a large negative value of $B_2^0$ for CeTa$_7$O$_{19}$ usually leads to the easy $c$-axis magnetic anisotropy, while a positive $B_2^0$ value indicates easy-plane magnetic anisotropy as in the case of Ce-based triangular antiferromagnets CeCd$_3$As$_3$\cite{Ce133} and CePtAl$_4$Ge$_2$\cite{Ce1142}. Furthermore, the large $B_4^3$ term implies a mixed ground state of $\left| \pm 1/2 \right\rangle $ and $\left| \mp 5/2 \right\rangle $. A small value of $B_4^0$ means the first excited CEF level would be a pure $\left| \pm 3/2 \right\rangle $ state. The expected consistent result has been demonstrated in CeTa$_7$O$_{19}$ and CeCd$_3$As$_3$\cite{Ce133}. For CePtAl$_4$Ge$_2$\cite{Ce1142}, its near zero $B_4^3$ term makes all CEF states being pure eigenstates. Meanwhile, the larger absolute value of $B_2^0$ lead to the larger magnetic anisotropy. Based on the CEF parameters, the anisotropy ratio of magnetic susceptibility for CeTa$_7$O$_{19}$ is $\chi_c$:$\chi_{ab}$$\sim$8.56 (T=1.8 K) using PyCrystalField software \cite{pycrystalfield}.

From the above CEF results, several conclusions about the magnetism of CeTa$_7$O$_{19}$ could be drawn. Firstly, the first excited CEF doublet lies at a very large energy gap 42.9~meV above the ground state, therefore CeTa$_7$O$_{19}$ can be considered an effective spin-1/2 system below room temperature. Secondly, the negative second-order CEF parameter means that CeTa$_7$O$_{19}$ possesses an easy-axis single-ion magnetic anisotropy along the $c$-axis. Thirdly, the knowledge on the CEF ground state allows us to determine the corresponding $g$-factor anisotropy using the equations below:

\begin{equation}
	\begin{aligned}
		g_{z}=2 g_{J}\left|\left\langle \pm \omega_{0}\left|J_{z}\right| \pm \omega_{0}\right\rangle\right|=2.57,\\
		g_{x y}=g_{J}\left|\left\langle \pm \omega_{0}\left|J_{ \pm}\right| \mp \omega_{0}\right\rangle\right|=0.86
	\end{aligned}
\end{equation} 
where $z$ and $xy$ denote directions along the $c$-axis and parallel to the $ab$-plane, respectively.

We further performed the electron spin resonance (ESR) measurements on CeTa$_7$O$_{19}$ to check the anisotropic $g$-factors. The well-resolved Ce ESR line is shown in Fig. 4. Through simulating the ESR spectra of CeTa$_7$O$_{19}$ using EasySpin\cite{EasySpin}, an open-source MATLAB toolbox, two g-factor eigenvalues with $g_1$=2.12 and $g_2$=0.96 could be obtained. Considering the calculated $g$-factors from CEF analysis, $g_1$ and $g_2$ correspond well with $g_z$ and $g_{xy}$ respectively, since these values are in fairly agreement besides a maximum difference of 18\% possibly caused by experimental errors.

\begin{figure}
	\includegraphics[width=7.5cm]{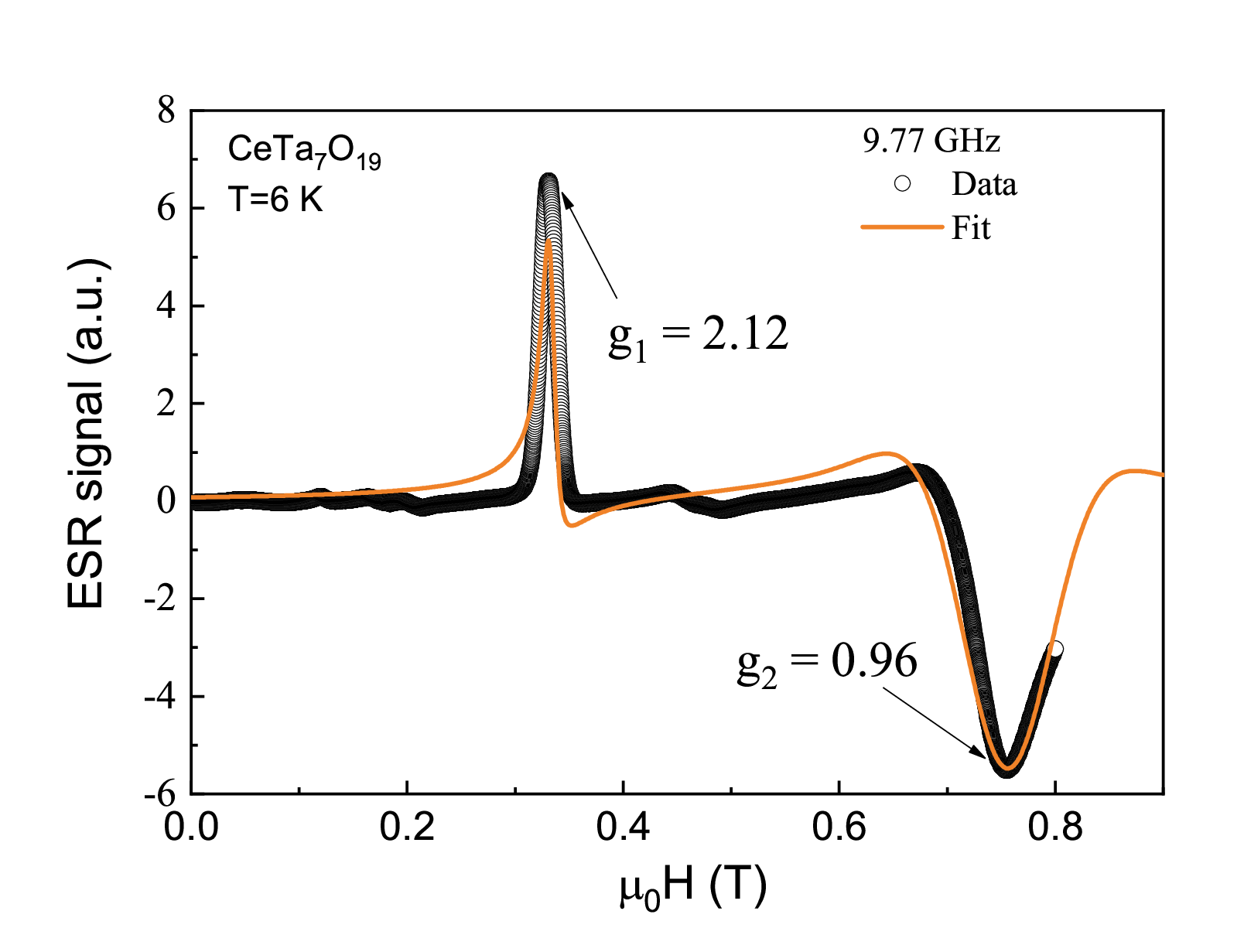}
	\caption {The ESR spectrum measured at 9.77~GHz and 6~K on CeTa$_7$O$_{19}$ powder samples. The solid line is the Lorentzian fit to the corresponding ESR signals.} \label{Fig4}
\end{figure}

\begin{figure}
	\includegraphics[width=7.4cm]{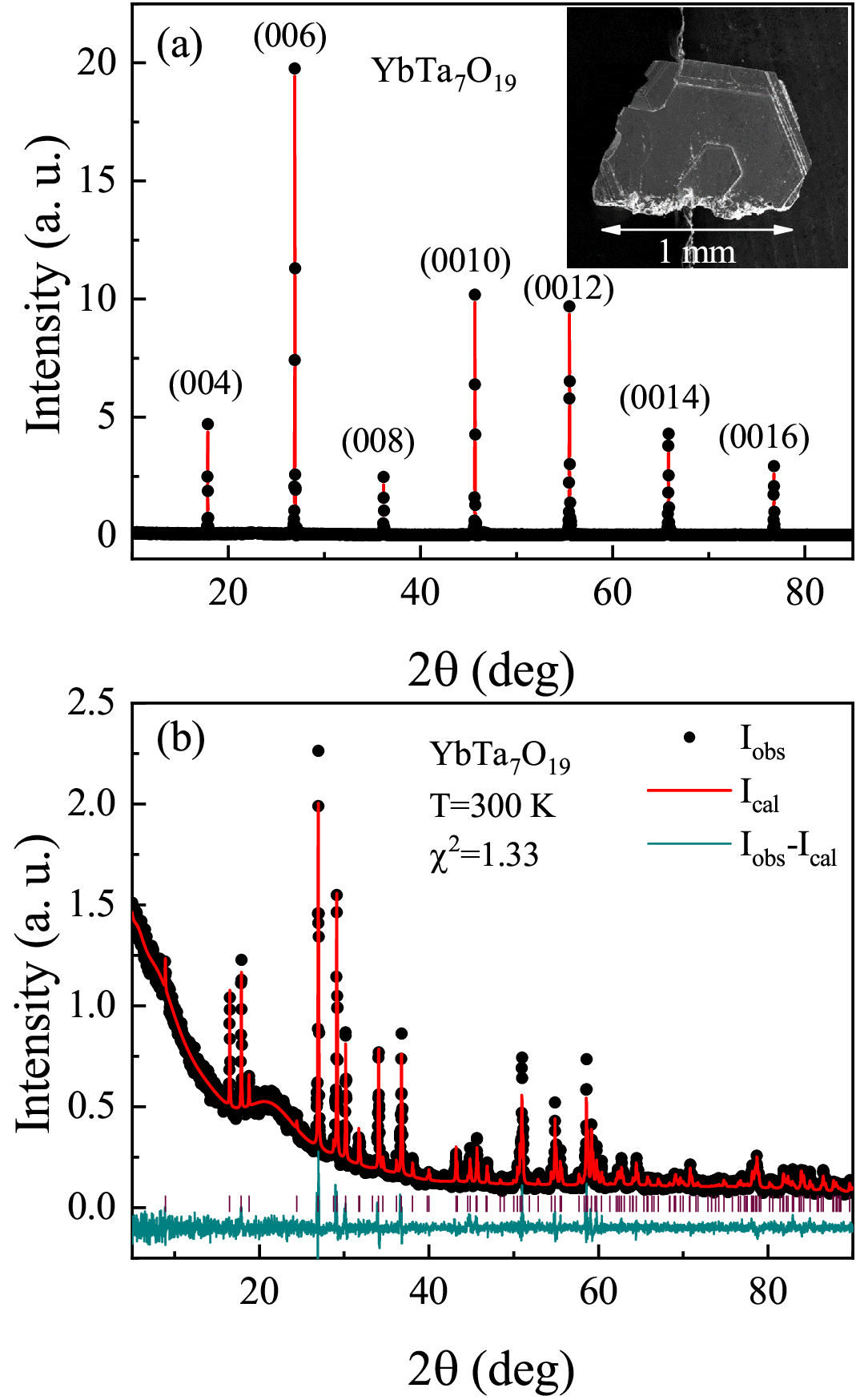}
	\caption {(a) The XRD patterns from the $ab$-plane of a YbTa$_7$O$_{19}$ single crystal. The inset shows a single crystal imaged by a scanning electron microscope. (b) XRD patterns of powders (crushed from single crystals) and the Rietveld refinement result.} \label{Fig5}
\end{figure}

The reliability of the CEF model obtained from the INS experiment could be further confirmed by simulating the magnetization data. As shown in Fig. 2(b) and (c), the simulated $\chi$(T) and M(H) curves based on non-interacting CEF model agree well with the experimental data. The deviation in the M(H) data at 1.8~K might be due to the exchange interaction which is not considered in the non-interacting CEF model. Furthermore, according to the $g$-factors obtained from CEF analysis, the $g$-factor of the powder sample should be $g_{pwd.}=\sqrt{g_z^2/3+2g_{xy}^2/3}=1.64$. Then based on the corresponding effective spin $J_{eff}$=1/2 model, the effective moment and saturation moment for CeTa$_7$O$_{19}$ powder sample should be $\mu_{eff}=\sqrt{J(J+1)}g_{pwd.}\mu_B=1.42~\mu_{B}$ and $\mu_{sat}=J g_{pwd.}\mu_B=0.82~\mu_{B}$. These two values achieve a good agreement with the effective moment derived from the CW fit on the low temperature susceptibility data [inset of Fig. 2(d)] and the saturation moment deduced from M(H) data at 1.8~K [Fig. 2(b)]. This consistency strongly suggests the CW temperature $\theta_{CW}$=-0.22~K obtained from the low temperature CW fit could also be a good estimation on the strength of antiferromagnetic interaction in CeTa$_7$O$_{19}$. Future higher resolution lower energy inelastic neutron scattering measurements could be used to probe fluctuations at this energy scale.

\begin{figure*}[htbp]
	\centering
	\includegraphics[width=0.85\textwidth]{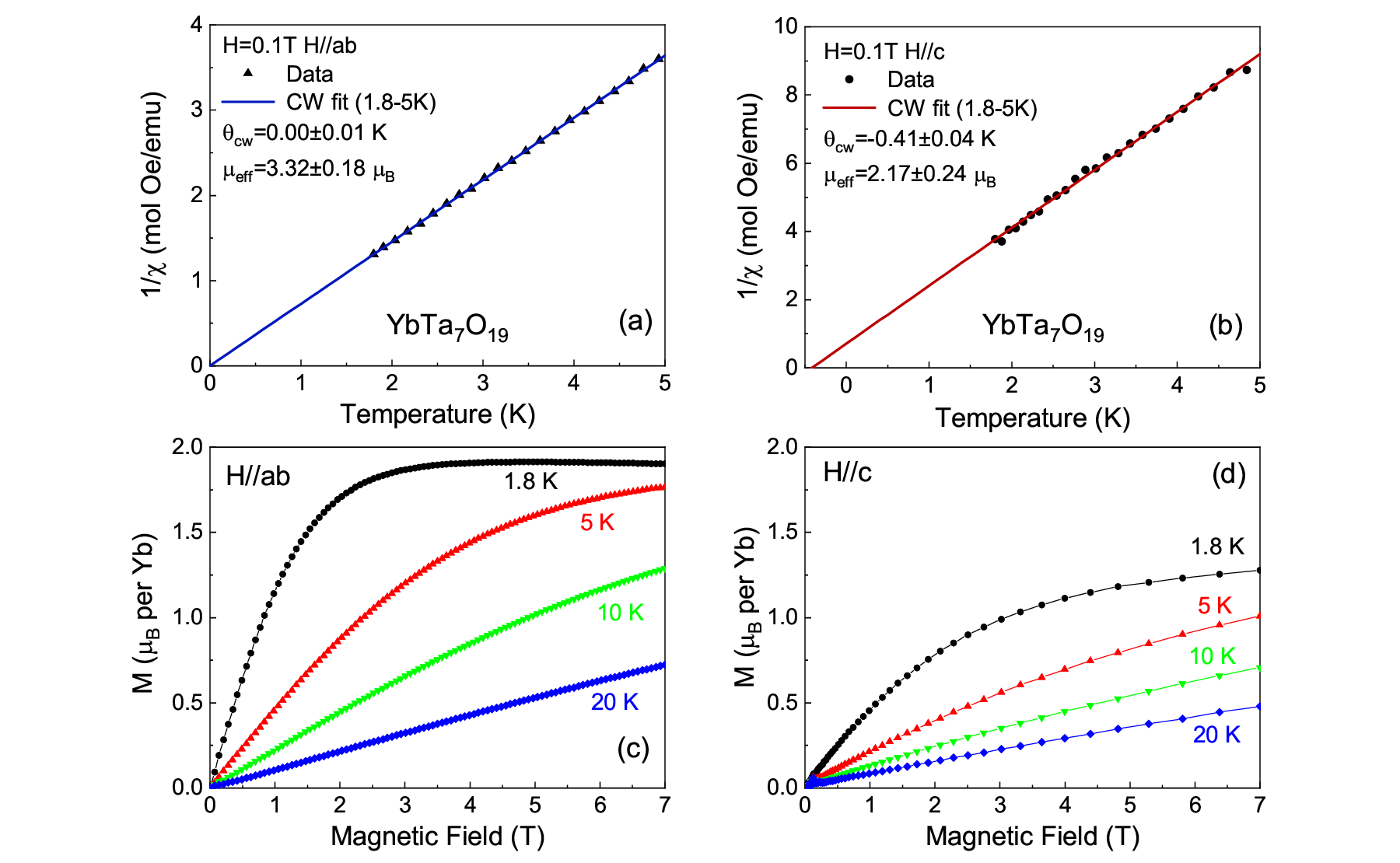}
	\caption {(a) The CW fit result on the low temperature magnetic susceptibility data under H$\parallel$ab. Similar fit result along H$\parallel$c is shown in (b). (c,d) Isothermal magnetization along H$\parallel$ab and H$\parallel$c at selected temperatures.} \label{Fig6}
\end{figure*}

\begin{figure}
	\includegraphics[width=7.5cm]{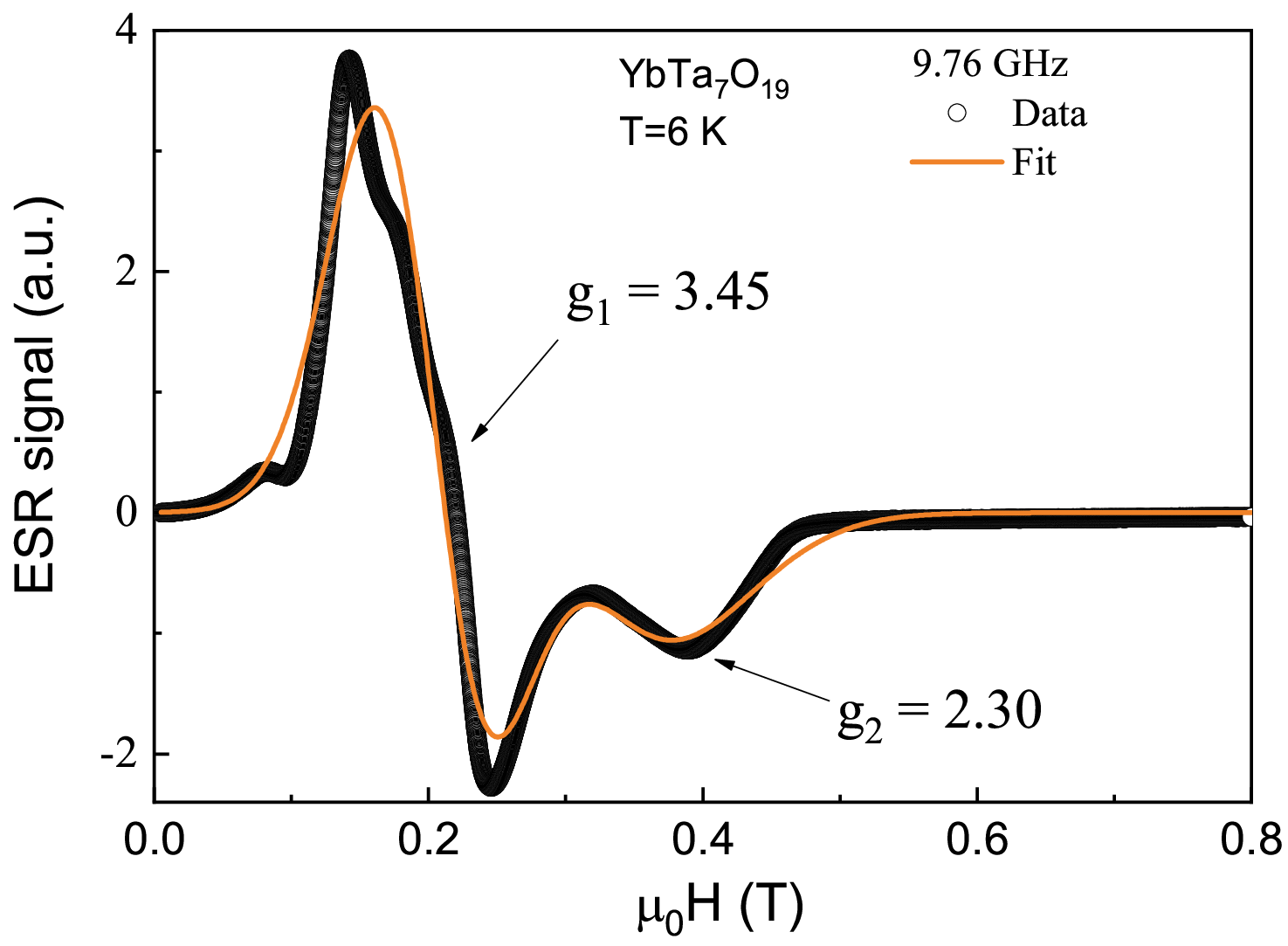}
	\caption {The ESR spectrum measured at 9.76~GHz and 6~K on YbTa$_7$O$_{19}$ powder samples. The solid line is the Lorentzian fit to the corresponding ESR signals.} \label{Fig7}
\end{figure}

Next, we could make a comparison of the magnetic properties between CeTa$_7$O$_{19}$ and previously reported NdTa$_7$O$_{19}$. Firstly the two compounds have a similar Ising-like $c$-axis magnetic anisotropy. Then based on the $g$-factors from CEF analysis and using the same calculation method considering an Ising-like spin Hamiltonian as in the report of NdTa$_7$O$_{19}$\cite{NM2022}, we could estimate the exchange anisotropy $J_z/J_{xy}=g_{z}^2/g_{xy}^2=9$, which is also highly anisotropic. Secondly, the estimated strength of antiferromagnetic exchange interaction in CeTa$_7$O$_{19}$ is 0.22~K which is only half of that in NdTa$_7$O$_{19}$ (0.46~K). Although the antiferromagnetic interaction is weak, theoretically, the perfect 2D triangular lattice and the exchange anisotropy close to the Ising limit in CeTa$_7$O$_{19}$ may also potentially serve to realize a QSL state\cite{NTOref10, NTOref11, NTOref12, NTOref16}. Further experimental characterization in ultra-low temperatures is needed for a final conclusion.

\subsection{YbTa$_7$O$_{19}$}

Based on some early reports on the single crystal growth of RETa$_7$O$_{19}$\cite{LnTa7O19-CVT, NdTa7O19-FLux, NdTa7O19-KMO, DyTa7O19-CVT}, we managed to grow YbTa$_7$O$_{19}$ single crystals with a modified condition as described in the METHODS section. As shown in the inset of Fig. 5(a), the single crystal of YbTa$_7$O$_{19}$ has a hexagonal shape and a typical 2D layered feature. From both the XRD on the $ab$-plane and the powder XRD refinement on the powders which are crushed from single crystals, YbTa$_7$O$_{19}$ could be confirmed to have the same crystal structure as CeTa$_7$O$_{19}$. The nearest Yb-Yb distance in the 2D triangular lattice is 6.202~\AA. Detailed crystallographic data are presented in the Supplemental Material\cite{supp}.

Next, we present the anisotropic magnetization data on the YbTa$_7$O$_{19}$ single crystal in Fig. 6. YbTa$_7$O$_{19}$ also does not show any sign of magnetic order down to 1.8~K. Similar as that of CeTa$_7$O$_{19}$ and NdTa$_7$O$_{19}$, a CW behavior of magnetic susceptibility is re-established at low temperatures below 5~K. Therefore, we use the CW model to fit the low temperature susceptibility data and the results are contrasting for magnetic field along $H\parallel ab$ ($\theta_{CW}$=0.0004~K, $\mu_{eff}$=3.32$\mu_{B}$) and $H\parallel c$ ($\theta_{CW}$=-0.41~K, $\mu_{eff}$=2.17$\mu_{B}$), as shown in Fig. 6(a) and (b). Furthermore, the anisotropic isothermal magnetization data directly reveal that YbTa$_7$O$_{19}$ has an easy $ab$-plane magnetic anisotropy with $\mu^{ab}_{sat.}\sim1.91\mu_{B}$ and $\mu^{c}_{sat.}\sim1.29\mu_{B}$ at 1.8~K, in contrast to $c$-axis anisotropy in CeTa$_7$O$_{19}$ and NdTa$_7$O$_{19}$, but similar as the anisotropy in YbMgGaO$_4$\cite{YMGO}.  Therefore a quantum XY-model may be expected for YbTa$_7$O$_{19}$ at low temperatures and possibly a topological phase transition following the so-called Berezinskii-Kosterlitz-Thouless scenario is also expected.

In order to obtain the anisotropic $g$-factors, the ESR measurement on YbTa$_7$O$_{19}$ powder was performed and the result is presented in Fig. 7. From the same fitting method described above, two $g$-factors with $g_1$=3.45 and $g_2$=2.30 are derived. Based on the magnetic anisotropy determined from magnetization, $g_1$ and $g_2$ should be $g_{xy}$ (parallel to the $ab$-plane) and $g_z$ (perpendicular to the $ab$-plane) respectively. Since Yb$^{3+}$ ion contains an odd number of electrons, the effective
spin may also be described by a ground state Kramers doublet. Based on this pesudospin-1/2 assumption and the $g$-factors from ESR, we could calculate the effective and saturated moment for YbTa$_7$O$_{19}$. The results are $\mu^{ab}_{eff}$=2.99$\mu_{B}$, $\mu^{c}_{eff}$=1.99$\mu_{B}$, $\mu^{ab}_{sat.}=1.73\mu_{B}$ and $\mu^{c}_{sat.}=1.15\mu_{B}$. These values are consistent with that obtained from magnetization data in Fig. 6, which suggests YbTa$_7$O$_{19}$ can also be considered an effective spin-1/2 system.

In Fig. 6(b), the CW temperature $\theta_{CW}$=-0.41~K by fitting $\chi_c$ suggests YbTa$_7$O$_{19}$ has an antiferromagnetic interaction strength comparable to NdTa$_7$O$_{19}$\cite{NM2022}. However, same fit on $\chi_{ab}$ yields a near zero value. The measurement on another crystal exhibits roughly the same contrasting values. The interpretation on this result may need the determination of CEF parameters through INS experiment, which is currently unavailable due to the lack of large amount of phase-pure YbTa$_7$O$_{19}$ powders. Additionally, the ESR line width of YbTa$_7$O$_{19}$ ($\sim$0.08~T) at 6~K is comparable with that of triangular antiferromagnets NaYbS$_2$\cite{ESR_Na} and KYbO$_2$\cite{ESR_K}, but much larger than that of CeTa$_7$O$_{19}$ ($\sim$0.02~T), which might suggest that the former compounds have stronger exchange interactions. However, there are many other factors which may broaden the ESR line width and not related to exchange interaction. Further ESR studies on the dilute magnetic doped crystal of RETa$_7$O$_{19}$ may give deep insights on the exchange interaction in this family of materials\cite{Zhangzheng}.

\section{Conclusions}

In summary, we have synthesized geometrically frustrated 2D TLAFs CeTa$_7$O$_{19}$ and YbTa$_7$O$_{19}$. We find that these materials exhibit no structural disorder. For CeTa$_7$O$_{19}$, the CEF excitations and parameters are determined from INS experiment. Combined with the magnetic susceptibility analysis, CeTa$_7$O$_{19}$ is identified as an effective spin-1/2 system with Ising-like magnetic anisotropy ($g_z$/$g_{xy}$$\sim$3) and weak antiferromagnetic exchange interaction ($J$$\sim$0.22~K). For YbTa$_7$O$_{19}$, its easy-plane magnetic anisotropy is directly identified by magnetization measurements on single crystal. Combined with ESR data analysis, the low temperature magnetic properties of YbTa$_7$O$_{19}$ agree well with the expectation from a ground state Kramer doublet with effective spin-1/2.

Although our initial structural and magnetic characterization of CeTa$_7$O$_{19}$
and YbTa$_7$O$_{19}$ shows that they could be considered as QSL candidate materials, it should be noted that their antiferromagnetic exchange interaction is much weaker than that of YbMgGaO$_4$ and AYbX$_2$, which usually have $J$ values of several kelvins. The reason should be the large distance between nearest RE-ions in the triangular lattice. As a result, it may be quite challenging for experimental characterizing the possible QSL state. On the other hand, the quite small $J$ in RETa$_7$O$_{19}$ means that their magnetic moments should be easily aligned by external magnetic field, causing a reduction of entropy. Therefore, both CeTa$_7$O$_{19}$ and YbTa$_7$O$_{19}$ could attract research interests as materials used in adiabatic demagnetization refrigeration\cite{ADR}.

\section*{Acknowledgement}
This work was supported by the National Natural Science Foundation of China  (No. 12074426, No. 12474148), the Fundamental Research Funds for the Central Universities, and the Research Funds of Renmin University of China (Grants No. 22XNKJ40). A portion of this research used resources at the Spallation Neutron Source, a DOE Office of Science User Facility operated by the Oak Ridge National Laboratory. The beam time was allocated to SEQUOIA spectrometer on proposal number IPTS-30317.1.

\bibliography{CeTaO}{}
\end{document}